\documentclass[runningheads]{llncs}
\usepackage{graphicx}
\usepackage{url}
\usepackage{color}
\usepackage{booktabs}
\usepackage[table,xcdraw]{xcolor}
\begin{document}
\title{Conditional Drums Generation using Compound Word Representations}
%


\author{Dimos Makris\inst{1} \and
Guo Zixun\inst{1} \and
Maximos Kaliakatsos-Papakostas\inst{2} \and
Dorien Herremans\inst{1}} 

\authorrunning{Dimos Makris et al.}


\institute{Information Systems Technology and Design, Singapore University of Technology and Design, Singapore\\
\email{\{dimosthenis\_makris,nicolas\_guo,dorien\_herremans\}@sutd.edu.sg}\\
\and
Institute for Language and Speech Processing, R.C. ``Athena'', Athens, Greece\\
\email{maximos@ilsp.gr}}

\maketitle              
\begin{abstract}
The field of automatic music composition has seen great progress in recent years, specifically with the invention of transformer-based architectures. When using any deep learning model which considers music as a sequence of events with multiple complex dependencies, the selection of a proper data representation is crucial. In this paper, we tackle the task of conditional drums generation using a novel data encoding scheme inspired by the Compound Word representation, a tokenization process of sequential data. Therefore, we present a sequence-to-sequence architecture where a Bidirectional Long short-term memory (BiLSTM) Encoder receives information about the conditioning parameters (i.e., accompanying tracks and musical attributes), while a Transformer-based Decoder with relative global attention produces the generated drum sequences. We conducted experiments to thoroughly compare the effectiveness of our method to several baselines. Quantitative evaluation shows that our model is able to generate drums sequences that have similar statistical distributions and characteristics to the training corpus. These features include syncopation, compression ratio, and symmetry among others. We also verified, through a listening test, that generated drum sequences sound pleasant, natural and coherent while they ``groove'' with the given accompaniment.

\end{abstract}
\section{Introduction}
Automatic music composition has slowly received more and more research attention over the last several decades~\cite{herremans2017functional,briot2020deep,deliege2006musical,herremans2017morpheus}. Many researchers have been approaching the task of music generation with a plethora of methods (e.g., ruled-based, grammars, probabilistic among others~\cite{papadopoulos1999ai,herremans2013composing}), recent research however, focuses intensely on deep generative architectures. The increased computational power available to us, along with  easier access to large musical datasets, can empower us to train models that generate much more realistic sounding music  (e.g.,~\cite{huang2018music}).

In this research, we focus on music generation in the symbolic domain. There are several diverse tasks in this domain, including chorale harmonisation~\cite{hadjeres2017deepbach}, piano~\cite{huang2020pop}, and multi-track generation~\cite{dong2018musegan} (see~\cite{briot2020deep,herremans2017functional} for further reading). An approach can be characterised as unconditional if it generates output from scratch, or \textbf{conditional} if it take additional input information to \emph{condition} the generation process, which can empower the user and making the generation process steerable. \textit{Conditioning} information may for instance be entire musical tracks (e.g., chord-conditioned melody generation~\cite{guo2021hrnn}) or specific constraints imposed by the user to control the generation output (e.g., emotions~\cite{makris2021generating,tan2020music} as well as general features such as musical style~\cite{papadopoulos2016assisted}).

In this work, we explore the task of conditional rhythm generation, specifically drums sequences. Although there have been numerous attempts at building multi-track generative systems, only few of them include drum tracks. In addition, existing research usually tackles the task in an unconditional manner with no accompanying tracks involved (see Section~\ref{sec:related} for examples). Therefore, we present a novel framework which utilizes a ``Encoder - Decoder'' scheme where a Bidirectional Long short-term memory (BiLSTM)~\cite{graves2005framewise} Encoder handles information for the conditioning parameters and a Transformer-based~\cite{vaswani2017attention} Decoder produces the generated drum sequences. Influenced by~\cite{randel1999harvard} and related work~\cite{makris2019conditional}, we use entire accompanying tracks as conditions, specifically Guitar and Bass along with extracted musical attributes such as Time Signature and Tempo.

The main contribution of this work lies in the encoding data representation. Transformer-based architectures are well known for their efficiency, which is achieved by applying attention mechanisms within large sequences of tokens. Since musical data is sequential, selecting the proper encoding representation is critical for such generative tasks. Hence, we propose a novel encoding scheme which is based on the \textbf{\underline{C}om\underline{p}ound Word} (CP) representation~\cite{hsiao2021compound} where musical events can be described in grouped tokens. We present different CP representations for the Encoder and Decoder in order to efficiently transcribe the events related to the accompanying (conditional) and target drum tracks.

The remainder of this paper is organised as follows: Section~\ref{sec:related} presents existing research on drums generation. Next, Section~\ref{sec:cp} \&~\ref{sec:arch} presents our proposed framework with a focus on the Encoding Representation and the utilised Architecture respectively. Sections~\ref{sec:exp_setup} \&~\ref{sec:exp_results} detail the experimental evaluation, while conclusions are presented in Section~\ref{sec:conclusions}.

\section{Related Work} \label{sec:related}
There has been extensive research on the task of drums generation, featuring different strategies and architectures. Although recent studies mainly use Neural Network-based architectures, there is some other notable work such as linear regression and K-Nearest Neighbors~\cite{wright2006towards} for generating expressive drum performance, or evolutionary algorithms~\cite{kaliakatsos2013evodrummer} to create rhythmic patterns by altering  a given base beat.

Another remarkable strategy was introduced by Choi et al.~\cite{cho2016music} who transcribed simultaneous drum events into  words and then used an LSTM with a seed sentence to generate drums sequences. Hutchings~\cite{hutchings2017talking} presented a framework based on a  \textit{sequence-to-sequence} (seq2seq) architecture which generates a track for  a full drum kit given a kick-drum sequence.

The work of Gillick et al.~\cite{gillick2019learning} can be considered as an important milestone, which introduced the first large scale symbolic dataset created by professional drummers. Their research focuses on producing expressive drums sequences for particular tasks such as tap2drum (generating a drum pattern based on an tapped rhythm by the user) and humanization (applying microtiming to quantised drums). Based on their dataset, Nuttall et al.~\cite{nuttall2021transformer} trained a Transformer-XL decoder for sequence generation and continuation.

In other work, Reinforcement Learning has been combined with neural network architectures. \textit{jaki}~\cite{brufordjaki}'s model is an example of such an approach. Their model generates 1-bar continuations of a seed sentence that can be controlled by the user through musical features such us density and syncopation. Karbasi et al.~\cite{karbasigenerative} proposed a model which learns rhythmic patterns from scratch through the interaction with other musical agents.

To best of our knowledge, there is very few existing research (see~\cite{lattner2019high} for the audio domain) that tackles the task of drums generation \emph{while considering accompanying tracks}. Although this can been addressed with multi-track generative systems, such as MuseGAN~\cite{dong2018musegan}, the closest model that inspired our research is the Conditional Neural Sequence Learners~\cite{makris2017combining,makris2019conditional} (CNSL). CNSL is a an architecture with stacked LSTM layers conditioned with a Feed-Forward layer to generate drum patterns. The Feed-Forward layer, which is called \textit{Conditional}, takes as input extracted features from the accompanying bass and guitar tracks along with other musical attributes such the metrical position within a bar. The model uses a piano-roll representation and obtained noteworthy performance considering the limited amount of training data.

\section{Data Encoding Representation}\label{sec:cp}
In this section we propose a novel data encoding scheme for the task of conditional drums generation, based on the \textbf{compound word} (CP) representation proposed by~\cite{hsiao2021compound}. CP is an alternative tokenization approach where multiple tokens that describe a specific event can be grouped into a ``super token'', or ``word''. It differs from the traditional token-based representations (i.e., MIDI-like~\cite{oore2020time} and the recent REMI~\cite{huang2020pop}) since musical pieces are not transcribed into a long $one$-dimensional stream of tokens. Instead, they are transcribed into smaller $n$-dimensional streams of tokens whereby $n$ is the number of grouped tokens that forms the length of a single CP word. This strategy improves the efficiency of Transformer-based~\cite{vaswani2017attention} architectures due to the decreased input sequence length which reduces the computational complexity. Recent studies show that CP achieves a better output quality compared to the aforementioned representations in certain tasks such as conditional/unconditional piano generation~\cite{hsiao2021compound,chou2021midibert} and emotion recognition~\cite{hung2021emopia}.

Our approach to generating a drum track conditioned with existing accompanying rhythm tracks (i.e., guitar and bass) motivates us to make use of the ``Encoder - Decoder'' scheme from the popular \textit{sequence-to-sequence} (seq2seq) architectures~\cite{cho2014learning}. Hence, we propose two different CP-based encoding representations whereby the Encoder handles information about the conditioning tracks, and the Decoder represents the generated drum sequences.

\begin{figure}[h]
\centering
\includegraphics[width=0.85\textwidth]{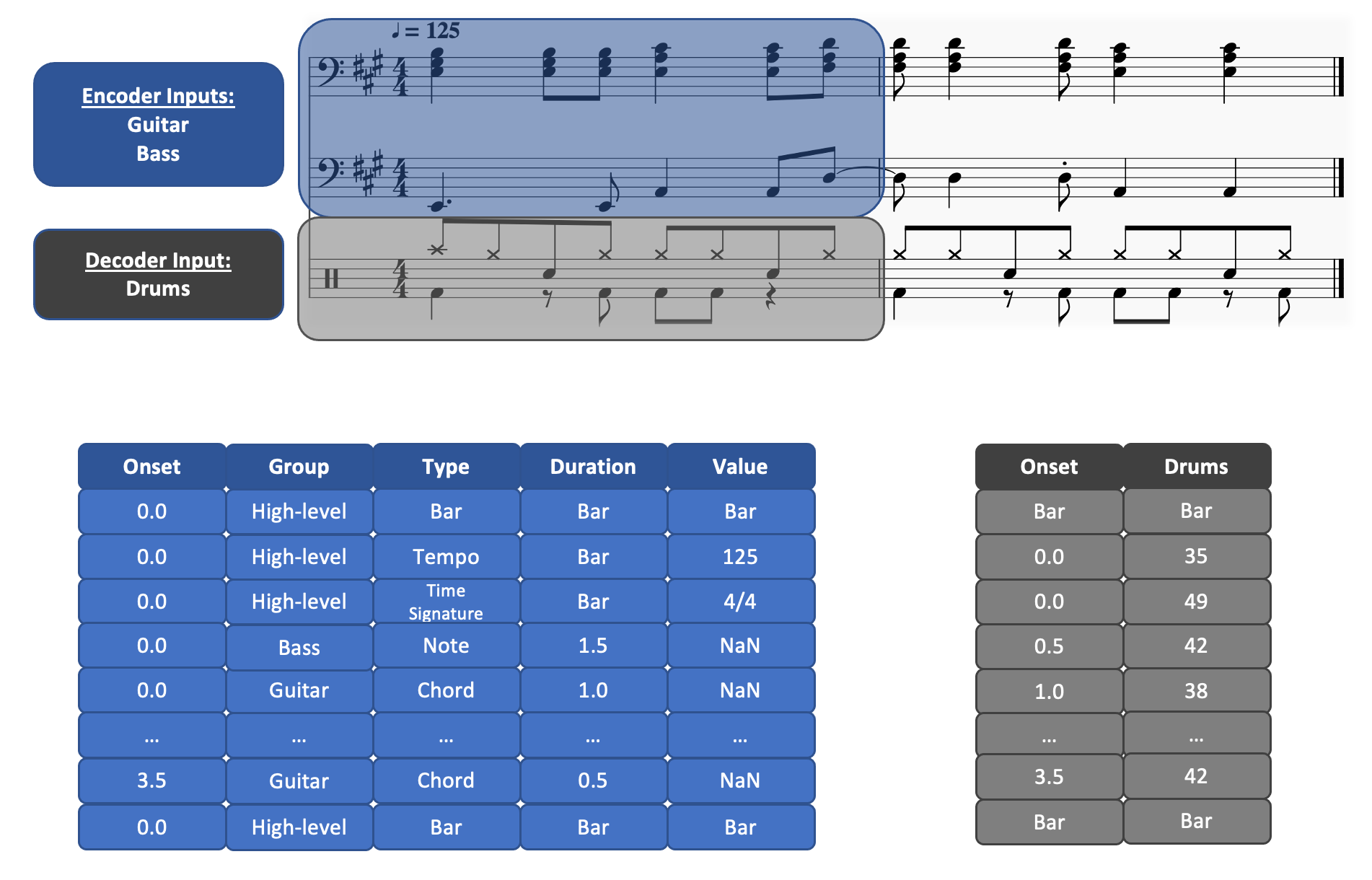}
\caption{Illustrated example of a training snippet represented in the proposed CP-based representation (for Encoder and Decoder inputs).}
\label{fig:encoding_rep}
\vspace{-22pt}
\end{figure}

\subsection{Encoder Representation - Conditional Information}
The idea to use guitar and bass tracks as the accompaniment input stems from fundamental musical principles of contemporary western music where the \textit{rhythm section}~\cite{randel1999harvard} typically consists of a drummer, a bass player, and a (at least one) chordal instrument player (e.g., guitarist). Therefore, drums are highly related to the bass and guitars tracks, perhaps more than any other instrument. Additionally, we take into account high-level parameters such as the tempo and time signature since it has been shown in related work~\cite{makris2017combining,makris2019conditional} that they can affect both the density and the ``complexity'' of the generated drum track.

For the Encoder we developed a 5-dimension CP representation in which every dimension corresponds to a specific one-hot encoded categories of tokens. Both the original CP and REMI are bar-based representations where time-slicing happens in each individual bar, we also adopt this. Additionally, we include high-level information that describes every bar in terms of Time Signature and Tempo. Thus, the resulting CP words can either describe \textbf{events} regarding the performance of the accompanying tracks, or specifying the values of these parameters. The proposed categories are the following (see Figure~\ref{fig:encoding_rep} for an example):

\begin{itemize}
    \item \textbf{Onset:} Describes the beginning of an event measured in quarter notes inside a bar. The maximum value is related to the current Time Signature of this bar. If a new bar occurs, it resets counting.
    \item \textbf{Group:} An event can either describe the performance of an accompanying instrument (i.e., Guitar or Bass) or specify a high-level parameter. The latter includes Tempo and Time Signature, as well as ``Bar'' event (this indicates the start of a new bar). In sum, there are three possible groups of events: Guitar, Bass or High-Level.
    \item \textbf{Type:} This category determines the content of the ``Group'' category. If it is marked as High-Level, it declares specifically which parameter is described (i.e., Tempo, Time Signature, or the starting of a new bar). On the other hand, if the event belongs to the Guitar or Bass group, it can be either a Note or Chord. 
    \item \textbf{Duration:} Indicates the duration of the event in quarter notes inside a bar. However, for high-level events the duration is marked with a ``Bar'' token.
    \item \textbf{Value:} Determines the value of the event, which depends on the declared Group and Type. Figure~\ref{fig:encoding_rep} shows the assigned values of each high-level parameter for the examined bar. Similar to~\cite{makris2017combining,makris2019conditional}, we exclude pitch information for Guitar and Bass events, since the drum pitches are not affected by the actual notes of accompanying tracks. Thus, \textit{NaN} tokens are used on these occasions.
\end{itemize}

\subsection{Decoder Representation - Generated Drum Sequences}

The Decoder outputs the generated drum events in the form of 2-dimension CP words. In addition, we make the assumption that drum notes do not have a duration and we exclude any information about velocities. Therefore the two categories that describe our representation are:

\begin{itemize}
    \item \textbf{Onset:} similar to the Encoder representation, the onset of a drum event is expressed in quarters inside a bar, and also includes a token to indicate the start of a new bar.
    \item \textbf{Drums:} the pitch value indicates which drum component is being played. In the case that multiple drum events happen at the same onset, then the new CP words have the same onset.
\end{itemize}

\section{Proposed Architecture}\label{sec:arch}
In this section we introduce the model architecture which includes our proposed encoding representation for the task of conditioned drums generation. Section~\ref{sec:4.1} reveals the components of the Encoder-Decoder architecture scheme, while Section~\ref{sec:4.2} reveals hyper-parameters and implementation details.

\subsection{Encoder - Decoder}\label{sec:4.1}
Our architecture consists of a BiLSTM~\cite{graves2005framewise} Encoder and a Transformer Decoder that uses Relative Global Attention~\cite{huang2018music}. Following~\cite{hsiao2021compound}'s methodology, our Encoder and Decoder inputs, which consist of the proposed CP representations, are fed to different Embedding spaces. The resulting latent variables are concatenated and fed as input to linear Dense layers. Therefore, these layers which include the combined embedding spaces of the CP words are acting as the ``actual'' inputs to our architecture.

\begin{figure}[htb]
\centering
\includegraphics[width=0.85\textwidth]{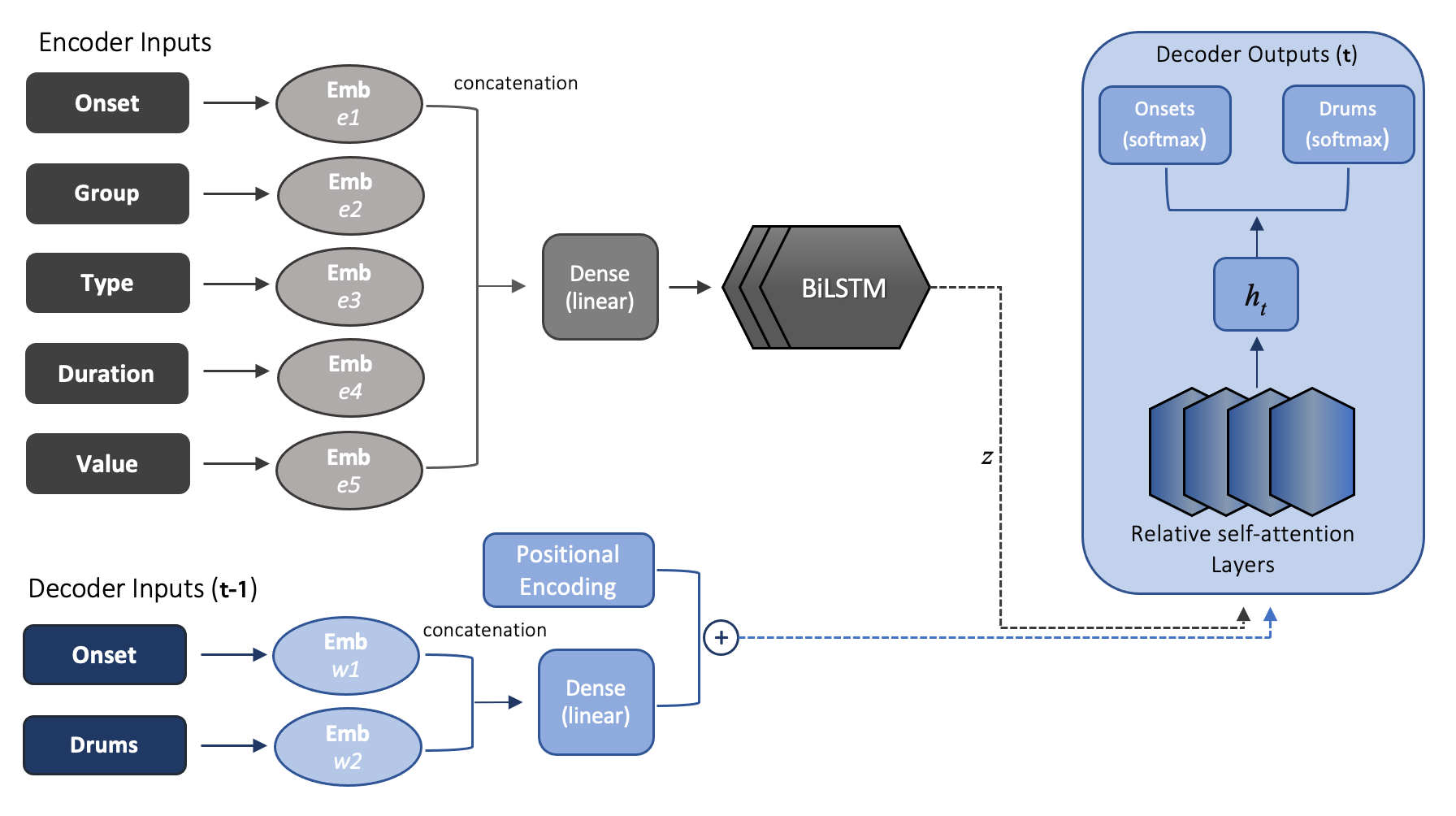}
\caption{Our proposed architecture features a stacked BiLSTM Encoder and a Music Transformer Decoder with Relative Global Attention~\cite{huang2018music}.}
\label{fig:arch}
\end{figure}

The resulting latent variable $z$ resulting from the Encoder, contains all of the information regarding the conditional tracks and the high-level features of an entire training piece. Accordingly, the Decoder, which is trained in an auto-regressive manner, inputs the projected embedding spaces at each time step $t$ - \textit{1} to the relative self attention layers, along with the latent variable $z$ from the Encoder. Since music has its underlying hierarchical structure (i.e. sections, phrases), musical events not only rely on the global position information but also the relative position of music events related to other music events. Hence, according to \cite{huang2018music}, music generated by self-attention layers with both relative and global position information is of better quality than that generated by vanilla self-attention layers~\cite{vaswani2017attention} with only the global position information. We therefore adopted \cite{huang2018music}'s efficient implementation of relative global attention to calculate the optimal size of the relative position information and set the relative attention window size to half of the total sequence length. The output logits from the relation-aware self-attention layers $h_t$ are projected to two separate dense layers followed by softmax activation. This way, we obtain the predicted onset and drum pitches from the shared hidden logits. Figure~\ref{fig:arch} shows the details of our proposed architecture for conditional drums generation.

\subsection{Implementation Details}\label{sec:4.2}

The Encoder consists of a 3-layer BiLSTM with 512 hidden units per layer, while the Transformer Decoder consists of a total of 4 self-attention layers with 8 multi-head attention modules. The number of hidden units in the Transformer's Feed-Forward layers was set to 1,024. We used the Tensorflow 2.x library~\cite{abadi2016tensorflow} and trained the model on a single Nvidia Tesla V100 GPU using Adam optimizer with a learning rate of $2e^{-5}$ and a weight decay rate of $0.01$. A 30\% dropout rate was set to all subsequent layers along with an early-stopping mechanism to prevent overfitting. 

Table~\ref{tab:emb} shows the size of the embedding layers for every input. These are directly correlated to the vocabulary size of the corresponding CP category (see Section~\ref{sec:dataset}). For the Decoder CP inputs, however, we slightly increased the embedding sizes compared to the Encoder, as we saw that this improved the performance. The final embedding sizes for the Encoder and Decoder CP words are 240 and 192 respectively. Finally, in order to enable diversity within the generation, we set a uniformly randomly sampling temperature ($\tau$) between 0.8 and 1.2 for both the Onset as well as Drums distribution output.

\begin{table}[htb]
\centering
\caption{Vocabulary and Embedding sizes for each CP category for the Encoder (Enc) and Decoder (Dec) input.}
\begin{tabular}{@{}l@{\hskip 0.4in}c@{\hskip 0.4in}c@{}}
\toprule
\textbf{CP-token Category} & \textbf{Voc. size} & \textbf{Emb. size} \\ \midrule
Onset (Enc)                & 31                 & 64                 \\
Group (Enc)           & 5                  & 16                 \\
Type (Enc)                 & 7                  & 32                 \\
Duration (Enc)             & 40                 & 64                 \\
Value (Enc)                & 33                 & 64                 \\ \midrule
Onset (Dec)                & 31                 & 96                 \\
Drums (Dec)                & 16                 & 96                 \\ \bottomrule
\end{tabular}
\label{tab:emb}
\vspace{-16pt}
\end{table}

\section{Experimental Setup}\label{sec:exp_setup}

The goal of our proposed framework is to generate drums sequences conditioned by given rhythm tracks (i.e., guitar and bass). Thus, a part of our experiments will focus on how to evaluate the quality of the generated drum rhythms. In addition, we compare the effectiveness of the proposed CP-based representation with existing approaches. In sum, we aim to evaluate the following: 
\begin{enumerate}
    \item Do the generated sequences have the same average densities (in terms of drum components) as the training dataset?
    \item Are the generated drum tracks of good musical quality, and how ``realistic'' do they sound in comparison to the corresponding ground-truth pieces from the test set for the given seeds?
    \item How effective is our proposed CP-based representation and architecture compared to existing approaches?
\end{enumerate}

We address these research questions by both calculating an extensive set of objective evaluation metrics as well as a listening study. Details of both of these are presented below.

\subsection{Dataset and Preprocessing}\label{sec:dataset}
For our experiments we use a subset of the LAKH MIDI dataset~\cite{raffel2016learning}. Specifically, we selected tracks that belong to the Rock and Metal music styles provided from the Tagtraum genre annotations~\cite{schreiber2015improving}. The reason for using these genres stems from the fact that the drum rhythms found in these genres are typically diverse and more complex compared to the contemporary popular music styles. After filtering out files that do not contain guitar, bass, and drum tracks we proceeded with the following preprocessing steps:
\begin{itemize}
    \item In case multiple tracks of either Drums, Bass, or Guitar occur within one song, we only take into account the track with the highest density of notes. It is worth mentioning that we select the aforementioned tracks based on information from their MIDI channel.
    \item We consider drum events with pitches that are common in the examined music styles. Therefore, our framework is able to produce the following drum components: Kick, Snare and Side-Stick, Open and Closed Hihats, 3 Toms, 2 Crashes, China, and finally, Ride Bell and Cymbals.
    \item In order to prevent long sequences, we break every training piece to phrases with a maximum length of 16 bars. According to our proposed representation, the maximum input lengths for the Encoder and Decoder are 597 and 545 CP words respectively.
    \item We eliminate phrases that contain rare Time Signatures and Tempos. The resulting training corpus contains phrases with 8 different Time Signatures (the most common is 4/4) and Tempos that vary from 60 to 220 BPM.
\end{itemize}

The resulting preprocessed dataset contains 14,176 phrases that were extracted from 2,121 MIDI files. This dataset was randomly divided into training / validation / test sets with a ratio of 8:1:1.

\subsection{Evaluation Metrics}

We conducted both a computational experiment as well as a user study to evaluate the quality of generated music and hence the effectiveness of the proposed method. Most of the recent studies~\cite{gillick2019learning,nuttall2021transformer} that focus on drums generation tasks rely completely on subjective evaluation using listening tests. Although there is no standard way to quantitatively measure whether generative models have been trained efficiently~\cite{agres2016evaluation}, we will include a quantitative evaluation as well as a listening study, by adapting metrics that have been used in similar tasks.

\subsubsection{Analytical Measures:}

We implemented two sets of metrics. Initially, we calculate the average number of specific drum components per bar. This is inspired by~\cite{dong2018musegan} and has been used widely in generative lead sheet systems (e.g., \cite{liu2018lead,makris2021generating}). Specifically we measure the density of Kick \& Snares, Hihats \& Rides, Toms, Cymbals along with the average percentage of empty bars. Although such calculations cannot guarantee to accurately measure the quality of generated music, they can reveal if the output generated by a model has pitch characteristics that resemble the learnt style (i.e. the training dataset).

On the other hand, the second set consists of metrics that offer one value for the entire generated drum sequence, and can be considered as a ``high-level'' feature set. Hence, it might indicate if generated pieces have a good rhythm, a long-term structure characteristics and if they ``groove'' with the given accompaniment. We adopt the following measures:

\begin{itemize}
    \item \textbf{Compression Ratio:} Reflects to the amount of repeated patterns that generated drums sequences contain. A high compression ratio is indicative of music with more structure and patterns, as per \cite{chuan2018modeling,guo2021hrnn}. We used the Omnisia\footnote{\url{https://github.com/chromamorph/omnisia-recursia-rrt-mml-2019}} library which implements the COSIATEC~\cite{meredith2013cosiatec} compression algorithm. 
    \item \textbf{Symmetry:} Indicates the repetitiveness of the (temporal) distance between consecutive onset events~\cite{kaliakatsos2012genetic}. Therefore, highly symmetric rhythms have equally spaced onset intervals.
    \item \textbf{Syncopation:} Measures displacements of the regular metrical accent in patterns. It is related to the Time Signature and refers to the concept of playing rhythms that accent or emphasize the offbeats. We used the normalised version of Longuet-Higgins and Lee's implementation~\cite{longuet1984rhythmic} adapted by~\cite{sioros2012measuring}.
    \item \textbf{Groove Consistency:} Calculates the similarity between each pair of grooving patterns across neighboring bars~\cite{wu2020jazz}. This metric takes into account the conditional tracks as well.
    \item \textbf{Pattern Rate:} Is defined as the ratio of the number of drum notes in a certain beat resolution to the total number of drum notes~\cite{dong2018musegan}.
\end{itemize}

\subsubsection{Listening Test Setup:}

We conducted an online listening test whereby participants rated short samples in order to evaluate the proposed framework. Each sample is in the form of a 3-track composition whereby the first two tracks are the accompanying or condition tracks, and the last one is the evaluated drum track. The drum tracks are either taken from the corresponding ground-truth sample of the test dataset, or generated by the newly proposed as well as baseline models. We asked the participants to rate the drum track of each sample on a 5-point Likert scale, ranging from 1 (very low) to 5 (very high), on the following criteria:

\begin{enumerate}
    \item Rhythm Correctness: Whether the overall rhythm is pleasant.
    \item Pitch Correctness: Is low if ``irregural'' drum notes are observed. This criteria differs from Rhythm Correctness as drums sequences may have a good rhythm while still containing unusual drum notes (i.e., playing only with toms).
    \item Naturalness: Does the drum track sound natural and ``human'' to the listeners?
    \item Grooviness: Do the drums ``groove'' with the given Guitar and Bass accompaniment? 
    \item Coherence: Whether the drums are consistent and contain repeated patterns.
\end{enumerate}

Inspired by similar listening tests for evaluating drums generation~\cite{gillick2019learning,nuttall2021transformer}, we asked the users to indicate if they thought the drum tracks are performed by human players (ground truth) or generated by an A.I. model (could be either the proposed model or the baseline model). We also added a third option, ``Not Sure'', in case the listener does not feel comfortable giving a direct answer. 

\section{Results}\label{sec:exp_results}

We compare our proposed method with state-of-the-art models that can generate conditional drums sequences. We also introduce different setups of our method by either changing the encoding representation or parts of the architecture. Specifically we implemented the following baseline models:

\begin{itemize}
    \item \textbf{CNSL:} As stated in Section~\ref{sec:related}, CNSL~\cite{makris2019conditional} is the closest related work for direct comparison. It differs from our approach since it uses \emph{piano-roll} representation as well as extracted features from the accompanying tracks as an input. We increased the size of the conditional window to two bars, and doubled the hidden sizes of LSTM and Dense layers for a fair comparison.
    \item \textbf{MuseGAN:} We adapted the conditional version of MuseGAN~\cite{dong2018musegan} to generate drum sequences only. Similar to CNSL, it uses a piano-roll representation. 
    \item \textbf{CNSL-Seq:} This is an altered \textit{seq2seq} version of the CNSL which uses the same architecture (i.e., conditional input as Encoder and drum output as Decoder) as our proposed method. As for the encoding representations, we use the original piano-roll representation for the Encoder, and our proposed CP representation for the Decoder (since Transformer-based Decoders are not compatible with piano-roll representations). The resulting encoding scheme can by characterised as a ``hybrid'' representation approach with a piano-roll Encoder and CP word-based Decoder. 
    \item \textbf{MT-CP:} We replaced the BiLSTM networks from our proposed architecture with relative self-attention layers in the Encoder stage, thus, resulting in the original implementation of Music Transformer~\cite{huang2018music}. This allows us to examine the impact of the Encoder type for conditional generation. 
\end{itemize}

\subsection{Objective Evaluation}

We generated a total of 1,418 files for every model, using input seeds from songs in our test set. The sampling hyper-parameter ($\tau$) was fixed to 1.0 for all cases. Table~\ref{tab:low_results} shows the results of the first set of objective metrics (i.e., low-level) representing the average density of different drum compenents per bar. Values close to those extracted from the training dataset may indicate that the generated fragments have a better chance to be musically valid as they match the density properties of existing music. Initially, we can observe that the CNSL and MuseGAN results are quite different than the real data. Our method performs best in three categories (Empty Bars, Toms, and Cymbals) with the CNSL-Seq version performing best in the other two categories (Kick-Snares and HH-Rides). In addition, fragments generated by the Music Transformer (MT-CP) version seem to be furthest away from the properties found in the training set, for in all categories of features.

\begin{table}[]
\centering
\caption{Results of the average density of drum components per bar. The values are better when they are closer to those computed from the training data.}
\begin{tabular}{@{}
>{\columncolor[HTML]{FFFFFF}}l@{\hskip 0.15in}  
>{\columncolor[HTML]{FFFFFF}}c@{\hskip 0.2in} 
>{\columncolor[HTML]{FFFFFF}}c@{\hskip 0.2in}  
>{\columncolor[HTML]{FFFFFF}}c@{\hskip 0.2in}  
>{\columncolor[HTML]{FFFFFF}}c@{\hskip 0.2in}  
>{\columncolor[HTML]{FFFFFF}}c @{}} \toprule
\textbf{} &
  \textbf{\begin{tabular}[c]{@{}c@{}}Empty \\ Bars\end{tabular}} &
  \textbf{\begin{tabular}[c]{@{}c@{}}Kick-\\ Snares\end{tabular}} &
  \textbf{\begin{tabular}[c]{@{}c@{}}HH-\\ Rides\end{tabular}} &
  \textbf{Toms} &
  \textbf{Cymbals} \\ \midrule
Training Dataset    & 5.91          & 5.0588          & 5.6203          & 0.7919          & 0.4902          \\ \midrule
\textbf{Ours} & \textbf{7.42} & 4.8703          & 5.8882          & \textbf{0.6075} & \textbf{0.3865} \\
CNSL & 10.24         & 4.7297          & 6.1037          & 0.9636          & 0.7552          \\
MuseGAN & 12.23         & 3.1254          & 4.2243          & 0.2216          & 0.1899  \\
CNSL-Seq    & 8.98      & \textbf{4.8795} & \textbf{5.5276} & 0.5342          & 0.3648  \\
MT-CP   & 10.45         & 4.0814          & 3.9925          & 0.4835          & 0.3531          \\
 \bottomrule       
\end{tabular}
\label{tab:low_results}
\vspace{-10pt}
\end{table}

Although our proposed model as well as the CNSL-Seq model both seem to achieve comparable average densities of drum components, both of which resemble the style of the training data, we cannot yet make conclusions about the quality of the models. In addition, we should keep in mind that a ``good'' drum sequence is mainly related to higher-level characteristics that are extracted from patterns instead of individual components. These types of features are examined in the second set of metrics. Considering that the ground-truth of the generated pieces is available since we used seeds from the test set, we can assess which model's output has a ``better resemblance'' to the ground-truth. This can be done by computing the absolute difference between the calculated values of each examined metric and the corresponding ground-truth value. Since we have over 1.4k generated pieces, we normalise these differences and estimate the mean of absolute differences. The resulting metric indicates ``how close''  the generated pieces are compared to the real piece for a particular high-level characteristic.

\begin{table}[]
\centering
\caption{Results for the high-level objective metrics: For each model and feature, we calculate the absolute difference between the extracted values for all generated pieces with the corresponding ground-truth values. Then, we compute the normalised mean of the aforementioned differences along with the standard deviation (in parentheses). Therefore, the lowest values are the best.}
\begin{tabular}{@{}
>{\columncolor[HTML]{FFFFFF}}l@{\hskip 0.03in}
>{\columncolor[HTML]{FFFFFF}}c@{\hskip 0.07in} 
>{\columncolor[HTML]{FFFFFF}}c@{\hskip 0.07in}  
>{\columncolor[HTML]{FFFFFF}}c@{\hskip 0.07in}  
>{\columncolor[HTML]{FFFFFF}}c@{\hskip 0.07in}  
>{\columncolor[HTML]{FFFFFF}}c@{}}
\toprule
\rowcolor[HTML]{FFFFFF} 
\textbf{} &
  \textbf{\begin{tabular}[c]{@{}c@{}}Compression\\ Ratio\end{tabular}} &
  \textbf{Symmetry} &
  \textbf{Syncopation} &
  \textbf{\begin{tabular}[c]{@{}c@{}}Groove\\ Consistency\end{tabular}} &
  \textbf{\begin{tabular}[c]{@{}c@{}}Pattern\\ Rate\end{tabular}} \\ \midrule
\rowcolor[HTML]{FFFFFF} 
\rowcolor[HTML]{FFFFFF} 
\cellcolor[HTML]{FFFFFF}\textbf{Ours}&
  \textbf{7.11} \textbf{(7.64)}&
  \textbf{7.55 (8.12)}&
  \textbf{3.76 (4.76)}&
  \textbf{1.36 (1.69)}&
  \textbf{1.54 (3.86)} \\
\rowcolor[HTML]{FFFFFF} 
CNSL       & 9.75 (7.87)  & 12.49 (11.45)         & 9.19 (8.82)          & 1.91 (2.17) & 3.16 ( 4.96) \\
\rowcolor[HTML]{FFFFFF} 
MuseGAN & 10.48 (9.55) & 15.45 (12.22)         & 14.54 (11.12)        & 5.12 (4.21) & 7.87 (10.14) \\
\rowcolor[HTML]{FFFFFF} 
CNSL-Seq    & 8.61 (8.14)  & 9.73 (10.52) & 7.07 (7.78) & 2.12 (2.07) & 3.88 (6.08)  \\
\rowcolor[HTML]{FFFFFF} 
MT-CP   & 8.99 (8.22)  & 10.51 (11.06)         & 8.44 (7.89)          & 2.01 (2.67) & 3.13 (5.34)  \\
 \bottomrule
\end{tabular}
\label{tab:high_results}
\end{table}

Table~\ref{tab:high_results} shows the results difference between the high-level features of generated pieces and those in the test set. Our proposed method has the lowest, thus the best, values compared to the baseline models for all feature categories. Especially for Symmetry and Syncopation we observe a huge difference compared to all the other models. Given that we are evaluating drum sequences, for which rhythm is arguably the most important feature, this is an excellent indicator of the success of our model. Similar to  Table~\ref{tab:low_results}, the worst results are obtained by the CNSL and conditional MuseGAN model. Since these model have a very similar task, this \textbf{confirms the effectiveness} of our proposed method. Once again the MT-CP model's results are significant worse than ours, which indicates that using a BiLSTM encoder instead of self-attention layers improves the overall performance. This has been confirmed in similar tasks, such us chord-conditioned melody generation~\cite{choi2021chord}.

Although our method, which features the CP-based representation, for the task of conditional drums generation achieves the best objective evaluation scores, the difference with the CNSL-Seq model is relative low. Considering the standard deviation of the high-level metrics, we cannot make strong conclusions about whether the hybrid approach with a piano-roll based Encoder representation is less efficient than ours. Therefore, we conducted a listening experiment as discussed in the next subsection.

\subsection{Subjective Evaluation}

For our listening experiment, we selected 30 tracks: 10 randomly generated by each model (ours and CNSL-Seq) and 10 randomly selected from the test set. These pieces were then rendered as YouTube videos with visualised piano-rolls. We believe that employing these visual cues can help participants to better perceive the drum patterns, in addition to listening to them. For each participant, 15 samples (out of these 30) were picked randomly during the listening study. A total of 41 subjects participated in our test with 74\% of them declaring medium or above musical background and knowledge. With a total of 615 votes, we get a high Cronbach's Alpha value (0.87) that confirms the consistency and reliability of the experiment. Table~\ref{tab:likert_results} reveals that our proposed model's ratings are significantly better compared to the pieces resulting from the CNSL-Seq model in all examined categories. Surprisingly, it even outperforms the real compositions when it comes to Naturalness, Grooviness and Coherence.

\begin{table}[htb]
\centering
\caption{Listening experiment ratings (Mean $\pm$ 95\% Confidence Interval) for drum sequences generated by our framework and the CNSL-Seq model (baseline), as well as existing ground-truth (real) compositions from the test set.}
\begin{tabular}{@{}
>{\columncolor[HTML]{FFFFFF}}l@{\hskip 0.05in}  
>{\columncolor[HTML]{FFFFFF}}c@{\hskip 0.12in}  
>{\columncolor[HTML]{FFFFFF}}c@{\hskip 0.12in}  
>{\columncolor[HTML]{FFFFFF}}c@{\hskip 0.12in}  
>{\columncolor[HTML]{FFFFFF}}c@{\hskip 0.12in}  
>{\columncolor[HTML]{FFFFFF}}c @{}}
\toprule
{\color[HTML]{000000} \textbf{}} &
  {\color[HTML]{000000} \textbf{\begin{tabular}[c]{@{}c@{}}Rhythm\\ Correctness\end{tabular}}} &
  {\color[HTML]{000000} \textbf{\begin{tabular}[c]{@{}c@{}}Pitch\\ Correctness\end{tabular}}} &
  {\color[HTML]{000000} \textbf{Naturalness}} &
  {\color[HTML]{000000} \textbf{Grooviness}} &
  {\color[HTML]{000000} \textbf{Coherence}} \\ \midrule
Real     & \textbf{3.59 (0.13)} & \textbf{3.42 (0.13)} & 3.13 (0.17) & 3.17 (0.16) & 3.32 (0.12) \\
CNSL-Seq & 3.34 (0.13) & 3.28 (0.14) & 3.06 (0.15) & 3.01 (0.15) & 3.18 (0.12) \\
\textbf{Ours}     & 3.56 (0.12) & 3.39 (0.14) & \textbf{3.34 (0.14)} & \textbf{3.31 (0.15)} & \textbf{3.39 (0.11)} \\ \bottomrule
\end{tabular}
\label{tab:likert_results}
\vspace{-12pt}
\end{table}

Table~\ref{tab:turing_results} shows the average accuracy of the participants' ability to predict if the drum rhythms were generated by an A.I. model or were human-composed. Interestingly, an average of 45.81\% of the human-composed (Real) sequences were mistaken as A.I., whereas only 39.02\% of the pieces generated by our proposed model were thought to be A.I. generated. This testifies to the perceived naturalness and quality of our model. On the other hand, the CNSL-Seq model performs significantly worse than the proposed, with 50.00\% rated as being A.I. generated and 21.94\% as human-composed. Thus, taking into account all of the aforementioned Likert ratings and users' predictions, our proposed model seems to generate drum sequences which sound more pleasant and natural compared to the baseline model. Samples from the experimental setup along with the model code and pre-processed dataset are available on GitHub\footnote{ \url{https://github.com/melkor169/CP_Drums_Generation}}.

\begin{table}[]
\vspace{-4pt}
\centering
\caption{Average Users' ratings to predict whether a drum sequence they listened to is generated by an A.I. model, is human-composed (Real), or they are not sure.}
\begin{tabular}{@{}
>{\columncolor[HTML]{FFFFFF}}l@{\hskip 0.2in} 
>{\columncolor[HTML]{FFFFFF}}c@{\hskip 0.2in} 
>{\columncolor[HTML]{FFFFFF}}c@{\hskip 0.2in} 
>{\columncolor[HTML]{FFFFFF}}c @{}}
\toprule
{\color[HTML]{000000} \textbf{Users' Predictions(\%)}} &
  {\color[HTML]{000000} \textbf{Real}} &
  {\color[HTML]{000000} \textbf{A.I.}} &
  {\color[HTML]{000000} \textbf{Not Sure}} \\ \midrule
Real               & 36.31                                  & \cellcolor[HTML]{FE0000}{\color[HTML]{000000} \textbf{45.81}} & 17.88 \\
CNSL-Seq (A.I.)      & 21.94                                  & \cellcolor[HTML]{32CB00}\textbf{50.00}                        & 28.06 \\
\textbf{Ours (A.I.)} & \cellcolor[HTML]{FE0000}\textbf{41.75} & 39.02                                                         & 19.23 \\ \bottomrule
\end{tabular}
\label{tab:turing_results}
\end{table}
\vspace{-14pt}

\section{Conclusions}\label{sec:conclusions}

This paper introduces a novel framework for conditional drums generation that takes into account the accompanying rhythm tracks (i.e., guitar and bass) and musical attributes (i.e., time signature and tempo). The proposed architecture consists of a BiLSTM Encoder which handles the conditioned input information and a Transformer-based Decoder that generates the drums sequences. The major contribution of this work relates to the proposed data representation which is based  on the compound word (CP) encoding scheme. CP-based representations have the advantage that an event can be described by multiple grouped tokens, thus leading to smaller sequence lengths  which reduces the computational complexity. Therefore, were present different CP representations for the Encoder and Decoder in order to tackle the task of conditional drums generation more efficiently.

The evaluate our proposed model and representation, we performed both analytical as well as more subjective experiments. During the analytical experiments, we compared variations of our method by changing the architecture and representation, with baselines from related work. The results show that our method has a solid high performance across the wide range of different quantitative metrics used for evaluation. Specifically, we show that our model is able to produce drums sequences conditioned to input tracks, with similar  pitch statistics to the training dataset, as well as high-level musical characteristics (i.e. syncopation, compression ratio etc.) compared to the corresponding ground truth test set. 

We also conducted a listening test to evaluate our proposed CP representation compared to a feature-based piano-roll representation  from the related work. The results show that our framework can generate drum sequences that sound more natural and pleasant. In addition, participants thought that our generated pieces were human, more so then actual human-generated pieces from the test set. As for future work, it would be interesting to examine the effect of adding more musical parameters and accompanying instruments to the Encoder representation.

\subsubsection{Acknowledgements} This work is supported by Singapore Ministry of Education Grant no. MOE2018-T2-2-161 and the Merlion PHC Campus France program.

\bibliographystyle{splncs04}
\bibliography{biblio}

\end{document}